\documentclass[graybox]{svmult}

\usepackage{mathptmx}       
\usepackage{helvet}         
\usepackage{courier}        
\usepackage{type1cm}        
\usepackage{makeidx}         
\usepackage{graphicx}        
\usepackage{multicol}        
\usepackage[bottom]{footmisc}
\usepackage{amsmath}
\usepackage{amssymb}

\makeindex             

\begin{document}


\title*{3D Higher spin gravity and the fractional quantum Hall effect}
\author{Mauricio Valenzuela}
\institute{Mauricio Valenzuela \at Facultad de Ingenier\'ia y Tecnolog\'ia,
Universidad San Sebasti\'an, General Lagos 1163, Valdivia 5110693, Chile. \email{mauricio.valenzuela@uss.cl}}
%
%
\maketitle

\abstract{This article is based on the talk ``Fractional Spin Gravity" presented in the 31\textit{st International Colloquium on Group Theoretical Methods in Physics}, Rio de Janeiro, 19-25th June 2016. There we emphasised an implication of the works \cite{Boulanger:2013naa,Boulanger:2015uha} by N. Boulanger, P. Sundell and the author on fractional spin extensions of $2+1D$ higher spin  gravity. This is that higher spin gravity may govern interactions of pseudo-particles excitations in the (fractional) quantum Hall effect. More generally, fractional spin currents in $2+1D$ source higher spin gravity curvatures.}

\section{Introduction: The fractional quantum Hall effect}

The fractional quantum Hall effect ({\bf FQHE}) occurs in a planar electric conductor when a perpendicular strong magnetic field  yields the condensation of an electron gas. The vacuum of this system consists of a quantum condensate of electrons trapped in Landau levels which form a new state of matter \cite{Tsui:1982yy}. The phases of the condensate are characterized by the value of the Landau level filling factor, $\nu$,  i.e. the number of Landau levels that are filled to the maximal capacity. The Hall conductance, $\sigma_H$, is induced by Lorentz forces on charges in the direction  perpendicular to the  current. Remarkably enough, $\sigma_H$ exhibits rational quantization. Its experimental values at FQHE conditions remain approximately constant, with respect to variations of the external magnetic field ($B$) in certain intervals, at rational number of times the fundamental unit of conductivity ($e^2/h$); these are the Hall plateaux. Thus at FQHE conditions the Hall conductance reads, $\sigma_H= \nu e^2/h$,  where $e$ is the electron charge, $h$ the Planck constant and $\nu\,\in \,\mathbb{Q}$ is a rational number. The integer quantum Hall effect refers to the cases of $\nu$ integer. It is well understood in terms of Landau level quantization without many-body electron interactions. It was predicted in \cite{Ando:1975} and then experimentally observed in \cite{vonKlitzing:1980pdk} for which Von Klitzing was awarded a Nobel Prize.   It was not expected however that the filling factors were quantized at values $\nu  <1$, until it was experimentally observed \cite{Tsui:1982yy}. A Nobel Prize went to Tsui and Stormer for this discovery, and to Laughlin for his explanation  \cite{Laughlin:1983fy}. $\nu=p/q$ represents the ratio of the number of electronic charges ($p$) to the number ($q=B/\phi_0$) of  units of magnetic flux quanta $\phi_0:=hc/e$. When the magnetic field varies an integer number of times the field quanta $\phi_0$ a phase transition must occur, meaning that some charges should either abandon a complete Landau level jumping to the next excited state, or, otherwise decaying to a lower Landau level. This is why the occurrence of values $\nu < 1$ are counterintuitive, as it means that there might be less than one unit of electric charge per magnetic field quanta, i.e. that exited states (and holes) have fractional charge. To explain this odd experimental fact the Coulomb many-body electron-electron interaction should be taken into account \cite{Laughlin:1983fy}.
Indeed, the Coulomb interaction of electrons lowers the gaps between Landau Levels,  breaking them  to fractions. As consequence the excitations of the many-electron problem appear to have fractional charge and angular momentum. From variational principles Laughling \cite{Laughlin:1983fy} was able to guess an ansatz for the wave function of that problem, which was actually very accurate. Laughlin imposed Wigner lattice symmetries on the many-electron configuration, as electrons repel each other, maximizing their distance inside a given Landau Level. The anzats for the ground states and excited states are dubbed ``Laughlin's" wave functions. He shown that they carry fractional charge and later on it was shown that they also carry fractional spin and statistics  \cite{Halperin:1984fn,Arovas:1984qr,Arovas:1985yb}. The excitated states behave as free particles, and they are known as ``anyons". Anyons have been observed experimentally \cite{dePicciotto:1997qc,Martin980,Willett:2010PRB,Willett02062009,vonKeyserlingk:2015yna}. Anyons are frequently presented as bound states of fractional charges and magnetic fluxes, i.e. magnetic instantons in the plane attached to charges (see e.g. \cite{Wilczek:1982wy,Arovas:1984qr,Itzhaki:2002rc}). The braided statistics of anyons can be interpreted as result of the braiding of magnetic fluxes attached  to the fractional charges when they rotate with respect to others.

The existence of anyons was predicted from theory before the experiments. In fact, based on the analysis of the topology of the configuration space of identical particles in $2+1$ dimensions and that the fundamental group of the rotation group in the spatial plane is $\mathbb{Z}$ (and not $\mathbb{Z}_2$ as for higher dimensions),  Leinaas and  Myrheim \cite{Leinaas:1977fm} argued in 1977 the existence of generalized braid statistics, generalizing bosonic and fermionic. Their article was considered an academic curiosity,  the existence of statistics different than bosonic and fermionic, in spite of their mathematical possibility, was too radical to be believed. Anyons appear also from group representation theory. Indeed, the Poincar\'e group, which is the isometry group of flat space-time, has also real-valued ($s \in \mathbb{R}$) spin representations in $2+1$ dimensions (for a review see \cite{Horvathy:2010vm}). Wigner's view on the existence of particles (fields) as carrying representations of the space-time isometries, together with Fradkin's belief on that ``all that is consistent is possible, and all that is possible happens" (in the 89's Dirac Medal  ceremony) are validated by the fractional quantum Hall effect: anyons do exist. 

The Hall setup can be imagined as a ``toy universe"  which passes through topological phase-transitions when the magnetic field (or the magnetic length $\ell_B= \sqrt{\hbar/eB}$ which is the typical scale at FQHE conditions) overpasses some critical values, but it stays stable for small variations. In particular in FQHE phases the anyons make their appearing playing the role of  fundamental particles. It is natural therefore to try to describe them by means of equations of motion, in the same way than Dirac's equation describes spin $1/2$ massive/massless fields. Some examples of anyon wave equations were given by Jackiw and Nair in \cite{Jackiw:1990ka}, and by Cortes and Plyushchay  in \cite{Cortes:1992fa}. The interested reader may consult reference \cite{Horvathy:2010vm} for a review of these and other anyon wave equations.  In \cite{Boulanger:2015uha} a topological (massless) first order non-linear action principle for  real-valued spin fields coupled to gauge gravity interactions was proposed, and which can be reduced to a Chern-Simons model. In the latter reference, one of the remarkable results is that the gravitational interaction of fractional spin fields are indeed of the higher spin ({\bf HS}) gravity type \cite{Blencowe:1988gj,Vasiliev:1989re}. It is therefore very suggestive that higher-spin gravity might describe the gravitational interactions of fractional spin fields at FQHE conditions. We shall argue below on theoretical grounds that gravitating anyons must couple to infinite-dimensional extensions of the Lorentz connection, which can be described using the tools of HS gravity \cite{Vasiliev:1992av,Prokushkin:1998bq,Vasiliev:1999ba,Vasiliev:2003ev}. Quoting Wigner's and Fradkin's ideas, we would expect that a systematic study of the interactions of anyons based on HS gravity should be consistent with their phenomenology in the FQHE. 

\section{Fractional higher spin gravity}

In an arbitrary $2+1D$ curved background the description of fractional spin fields requires the introduction of a Lorentz connection taking values in a discrete serie representation \cite{Bargmann:1946me,Barut:1965}, which are infinite dimensional and so the fractional spin fields must have also infinite components. The formulation of a Chern-Simons action principle, which make uses of traces of product of Lorentz generators, might be inconsistent when (infinite) matrix representations are used since traces may either diverge or converge to wrong (Lorentz symmetry broken) values. These problems can be fixed using Vasiliev's HS gravity technology using regularized (super) trace definitions  (see e.g. \cite{Vasiliev:1999ba}). Doing so we were able  in reference \cite{Boulanger:2015uha} to write down an action principle for fractional spin gravity. As we see, a technical need suggests to look at HS gravity in order to construct a consistent framework of gauge gravity interactions of fractional spin fields. From another point of view, since fractional spin fields must transform under an infinite dimensional (discrete) representations of the Lorentz algebra they will also admit the action of the universal enveloping algebra of the Lorentz algebra. The latter algebra defines also the gauge algebra of HS Chern-Simons gravity  \cite{Blencowe:1988gj,Vasiliev:1989re}, therefore it becomes natural to promote the whole HS algebra, not only the Lorentz generators, to gauge fields mediating also interactions of fractional spin fields. Doing so the spin-$2$ gravity interactions of fractional spin fields are extended by infinite many fields with arbitrary (half-)integer spins. Thus, as pointed out, HS gravity provides natural interactions for fractional spin fields. Let us write down the equations of motion of the fractional spin gravity theory \cite{Boulanger:2013naa,Boulanger:2015uha} in order to observe this more explicitly. These are:  
 \begin{eqnarray}
&& {\rm d}W+ W \, W +\Psi  \, \overline{\Psi}=0  \,, \qquad {\rm d}U+U \, U + \overline{\Psi}  \, \Psi  =0  \, , \label{eq1}\\
&& {\rm d}\Psi + W \,\Psi  + \Psi \,U =0  \, , \qquad {\rm d}\overline{\Psi} + \overline{\Psi}  \,W  + U \, \overline{\Psi}=0  \,,\label{eq2}
\end{eqnarray}
which are more succinctly expressed as the vanishing of the gauge curvature,
\begin{equation}\label{A}
{\rm d}\mathbb{A}+ \mathbb{A}^2 = 0\,, \qquad  \mathbb{A}=\left[
\begin{array}{cc}
W & \Psi \\
\overline{\Psi} & U
\end{array}
\right] \,,
\end{equation}
where $\mathbb{A}$ contains the fusion of HS gravity connection ($W$), a fractional spin one forms ($\psi$), and a $U \in U(\infty)\otimes U(\infty)$ non-abelian internal field.  As shown in \cite{Boulanger:2013naa,Boulanger:2015uha} the $\psi$ fields are valued in non-polynomial class of functions of universal enveloping algebra generators. Thus the density $\Psi  \, \overline{\Psi}$, when they are Taylor expanded, sources the field strength (${\rm d}W+ W \, W$) of HS gravity for all spins. The same is valid for the non-abelian curvature ${\rm d}U+ U \, U$. Thus non-trivial anyons distributions source higher-spin gravity and non-abelian interactions. When $\Psi=0=U$ the system \eqref{eq1} is equivalent to the Chern-Simons HS gravity  \cite{Blencowe:1988gj,Vasiliev:1989re}. For definiteness and simplicity let us choose $W$ valued in a representation of the HS algebra  $Aq^e_+(2;\mu)$  \cite{Vasiliev:1989re}, up to tensor-product extensions. $W$ has an expansion of the type
\begin{equation}
W= \sum_{a=0,1,2; n=0,1,...,\infty} dx^I \frac{1}{n!} W_I ^{a_1a_2...a_n} J_{a_1a_2...a_n},
\end{equation}
up to supersymmetric extensions by fermionic (spinor) components, and idempotent generators. $dx^I$ are line elements and  $J_{a_1a_2...a_n}$ are symmetric tensors belonging to the universal enveloping algebra of the Lorentz algebra generated by elements $J_a$. Hence $J_{a_1a_2...a_n}=J_{(a_1}\cdots J_{a_n)}$ consists of symmetric products with spin $n$. 
The parameter $\mu$ in $Aq^e_+(2;\mu)$ determines the lowest spin of the anyons \cite{Boulanger:2013naa,Boulanger:2015uha}, 
\begin{equation}\label{s}
s=\tfrac{1+\mu}{4}.
\end{equation}
 It was noticed in \cite{Boulanger:2013naa,Boulanger:2015uha} that for critical values, $\mu=-(2\ell +1),\, \ell=0,1,2,...$,  anyons become bosons or  fermions since $s=-\tfrac{\ell}{2}$, and the $Aq^e_+(2;\nu)$  algebra is truncated to a matrix algebra $Mat_{\ell+1}(\mathbb{C})$ \cite{Vasiliev:1989re}, while the $u(\infty)$ algebra is truncated to $u(\ell)$. With suitable reality conditions, $W$ and $U$ will take values respectively in the algebras $sl( \ell +1, \mathbb{R})$ and $u(\ell)$ (up to tensor products),  while the boson/fermion (before anyon) $\Psi$ and $\bar{\Psi}$ transform under the one-sided action  of these algebras (cf.  \eqref{eq1}-\eqref{eq2}) of spin $s$.  
Thus the model \eqref{eq1} contains $SL(\ell)$-type of HS gravities. 

Though HS gravity interactions in a setup such like the Hall effect might be expected to be weak, because their topological nature it does not mean trivial. Indeed the braided statistics of anyons is yield by Chern-Simons interactions \cite{Arovas:1984qr}. Thus, even if  HS gravity fields do not propagate, non-trivial topological configurations might be reflected in bulk-edge effects, as the Hall conductance for instance. To find out what possible effects from HS gravity can be measured  needs however further research. Our goal here has been to point out that HS gravity may be predictive in the FQHE and similar experiments.

\section{Conclusions}
We have argued that the effects of higher spin gravity \cite{Boulanger:2013naa,Boulanger:2015uha} may be observed in the FQHE and similar experiments. Consider for instance the statistical phases of anyons in the Quantum Hall effect, given in terms of the filling factor by $\exp(i\pi \nu)$, and the statistical phases $\exp(-i2\pi s)$ (see \eqref{s}) in the models \cite{Boulanger:2013naa,Boulanger:2015uha}. Comparing both statistical phases we obtain that the filling factor of the FQHE
\begin{equation}\label{nus}
\nu=2s=\tfrac{1+\mu}{4},
\end{equation}
is related to the $\mu$-parameter of the fractional spin algebra \eqref{eq1}-\eqref{eq2}. 
On the one hand, for critical values, $\mu=-(2\ell+1)$, the  model \cite{Boulanger:2013naa,Boulanger:2015uha} is reduced to $SL(\ell) \otimes SL(\ell)$ (matrix) Chern-Simons HS gravity and  the spin of the fractional spin fields become bosonic/fermionic $|s|=\tfrac{\ell}{2}$.  
Thus $SL(\ell) \otimes SL(\ell)$ HS gravities may be related to gravity interactions of the fundamental excitations/holes in the \textit{integer quantum Hall effect} ($ \nu= \ell$). More generally,  for non-critical values of the HS algebra parameter $\mu$, and the related non-integer filling factor \eqref{nus}, we would expect that their interactions will be described by the fractional spin gravity model \eqref{eq1}-\eqref{eq2} or extensions of it. 
In this way, we expect that fractional spin gravity \cite{Boulanger:2013naa,Boulanger:2015uha} will contribute to a better comprehension of the FQHE and related phenomena (e.g.  \cite{Laughlin:1988fs,Chen:1989xs,Fetter:1991ju}).

{\bf Acknowledgements:} I would like to thank to N. Boulanger and P. Sundell for their enlightening comments and their collaboration in this project.

\providecommand{\href}[2]{#2}\begingroup\raggedright 


\begin{thebibliography}{99.}%
\bibitem{Boulanger:2013naa}
N.~Boulanger, P.~Sundell, and M.~Valenzuela, ``{Three-dimensional
  fractional-spin gravity},'' {\em JHEP} {\bf 02} (2014) 052,
  \href{http://arXiv.org/abs/1312.5700}{{\tt 1312.5700}}.
[Erratum: JHEP03,076(2016)].

\bibitem{Boulanger:2015uha}
N.~Boulanger, P.~Sundell, and M.~Valenzuela, ``{Gravitational and gauge
  couplings in Chern-Simons fractional spin gravity},'' {\em JHEP} {\bf 01}
  (2016) 173, \href{http://arXiv.org/abs/1504.04286}{{\tt 1504.04286}}.
[Erratum: JHEP03,075(2016)].

\bibitem{Tsui:1982yy}
D.~C. Tsui, H.~L. Stormer, and A.~C. Gossard, ``{Two-dimensional
  magnetotransport in the extreme quantum limit},'' {\em Phys. Rev. Lett.} {\bf
  48} (1982)
1559--1562.

\bibitem{Ando:1975}
T.~Ando, Y.~Matsumoto, and Y.~Uemura, ``Theory of hall effect in a
  two-dimensional electron system,'' {\em Journal of the Physical Society of
  Japan} {\bf 39} (1975), no.~2, 279--288.

\bibitem{vonKlitzing:1980pdk}
K.~von Klitzing, G.~Dorda, and M.~Pepper, ``{New method for high accuracy
  determination of the fine structure constant based on quantized Hall
  resistance},'' {\em Phys. Rev. Lett.} {\bf 45} (1980)
494--497.

\bibitem{Laughlin:1983fy}
R.~B. Laughlin, ``{Anomalous quantum Hall effect: An Incompressible quantum
  fluid with fractionallycharged excitations},'' {\em Phys. Rev. Lett.} {\bf
  50} (1983)
1395.

\bibitem{Halperin:1984fn}
B.~I. Halperin, ``{Statistics of quasiparticles and the hierarchy of fractional
  quantized Hall states},'' {\em Phys. Rev. Lett.} {\bf 52} (1984) 1583--1586.
[Erratum: Phys. Rev. Lett.52,2390(1984)].

\bibitem{Arovas:1984qr}
D.~Arovas, J.~R. Schrieffer, and F.~Wilczek, ``{Fractional Statistics and the
  Quantum Hall Effect},'' {\em Phys. Rev. Lett.} {\bf 53} (1984)
722--723.

\bibitem{Arovas:1985yb}
D.~P. Arovas, J.~R. Schrieffer, F.~Wilczek, and A.~Zee, ``{Statistical
  Mechanics of Anyons},'' {\em Nucl. Phys.} {\bf B251} (1985)
117--126.

\bibitem{dePicciotto:1997qc}
R.~de~Picciotto, M.~Reznikov, M.~Heiblum, V.~Umansky, G.~Bunin, and D.~Mahalu,
  ``{Direct observation of a fractional charge},'' {\em Nature} {\bf 389}
  (1997)
162--164.

\bibitem{Martin980}
J.~Martin, S.~Ilani, B.~Verdene, J.~Smet, V.~Umansky, D.~Mahalu, D.~Schuh,
  G.~Abstreiter, and A.~Yacoby, ``Localization of fractionally charged
  quasi-particles,'' {\em Science} {\bf 305} (2004), no.~5686, 980--983.

\bibitem{Willett:2010PRB}
R.~L. Willett, L.~N. Pfeiffer, and K.~W. West, ``Alternation and interchange of
  e/4 and e/2 period interference oscillations consistent with filling factor
  5/2 non-abelian quasiparticles,'' {\em Phys. Rev. B} {\bf 82} (Nov, 2010)
  205301.

\bibitem{Willett02062009}
R.~L. Willett, L.~N. Pfeiffer, and K.~W. West, ``Measurement of filling factor
  5/2 quasiparticle interference with observation of charge e/4 and e/2 period
  oscillations,'' {\em Proceedings of the National Academy of Sciences} {\bf
  106} (2009), no.~22, 8853--8858.

\bibitem{vonKeyserlingk:2015yna}
C.~W. von Keyserlingk, S.~H. Simon, and B.~Rosenow, ``{Enhanced Bulk-Edge
  Coulomb Coupling in Fractional Fabry-Perot Interferometers},'' {\em Phys.
  Rev. Lett.} {\bf 115} (2015), no.~12,
126807.

\bibitem{Wilczek:1982wy}
F.~Wilczek, ``{Quantum Mechanics of Fractional Spin Particles},'' {\em
  Phys.Rev.Lett.} {\bf 49} (1982)
957.

\bibitem{Itzhaki:2002rc}
N.~Itzhaki, ``{Anyons, 't Hooft loops and a generalized connection in
  three-dimensions},'' {\em Phys. Rev.} {\bf D67} (2003) 065008,
\href{http://arXiv.org/abs/hep-th/0211140}{{\tt hep-th/0211140}}.

\bibitem{Leinaas:1977fm}
J.~Leinaas and J.~Myrheim, ``{On the theory of identical particles},'' {\em
  Nuovo Cim.} {\bf B37} (1977)
1--23.

\bibitem{Horvathy:2010vm}
P.~A. Horvathy, M.~S. Plyushchay, and M.~Valenzuela, ``{Bosons, fermions and
  anyons in the plane, and supersymmetry},'' {\em Annals Phys.} {\bf 325}
  (2010) 1931--1975,
\href{http://arXiv.org/abs/1001.0274}{{\tt 1001.0274}}.

\bibitem{Jackiw:1990ka}
R.~Jackiw and V.~Nair, ``{Relativistic wave equations for anyons},'' {\em
  Phys.Rev.} {\bf D43} (1991)
1933--1942.

\bibitem{Cortes:1992fa}
J.~Cortes and M.~Plyushchay, ``{Linear differential equations for a fractional
  spin field},'' {\em J.Math.Phys.} {\bf 35} (1994) 6049--6057,
\href{http://arXiv.org/abs/hep-th/9405193}{{\tt hep-th/9405193}}.

\bibitem{Blencowe:1988gj}
M.~Blencowe, ``{A Consistent Interacting Massless Higher Spin Field Theory in
  $D$ = (2+1)},'' {\em Class.Quant.Grav.} {\bf 6} (1989)
443.

\bibitem{Vasiliev:1989re}
M.~A. Vasiliev, ``Higher spin algebras and quantization on the sphere and
  hyperboloid,'' {\em Int.J.Mod.Phys.} {\bf A6} (1991) 1115--1135.

\bibitem{Vasiliev:1992av}
M.~A. Vasiliev, ``{More on equations of motion for interacting massless fields
  of all spins in (3+1)-dimensions},'' {\em Phys. Lett.} {\bf B285} (1992)
225--234.

\bibitem{Prokushkin:1998bq}
S.~F. Prokushkin and M.~A. Vasiliev, ``{Higher-spin gauge interactions for
  massive matter fields in 3D AdS space-time},'' {\em Nucl. Phys.} {\bf B545}
  (1999) 385,
\href{http://arXiv.org/abs/hep-th/9806236}{{\tt hep-th/9806236}}.

\bibitem{Vasiliev:1999ba}
M.~A. Vasiliev, ``{Higher spin gauge theories: Star-product and AdS space},''
\href{http://arXiv.org/abs/hep-th/9910096}{{\tt hep-th/9910096}}.

\bibitem{Vasiliev:2003ev}
M.~A. Vasiliev, ``{Nonlinear equations for symmetric massless higher spin
  fields in (A)dS(d)},'' {\em Phys. Lett.} {\bf B567} (2003) 139--151,
\href{http://arXiv.org/abs/hep-th/0304049}{{\tt hep-th/0304049}}.

\bibitem{Bargmann:1946me}
V.~Bargmann, ``{Irreducible unitary representations of the Lorentz group},''
  {\em Annals Math.} {\bf 48} (1947)
568--640.

\bibitem{Barut:1965}
A.~O. Barut and C.~Fronsdal, ``On non-compact groups, ii. representations of
  the 2+1 lorentz group,'' {\em Proc. Roy. Soc. London} {\bf A287} (1965)
  532--548.

\bibitem{Boulanger:2013zla}
N.~Boulanger, P.~Sundell, and M.~Valenzuela, ``{A Higher-Spin Chern-Simons
  Theory of Anyons},'' {\em Phys. Part. Nucl. Lett.} {\bf 11} (2014), no.~7,
  977--980, \href{http://arXiv.org/abs/1311.4589}{{\tt 1311.4589}}.
[Erratum: Phys. Part. Nucl. Lett.13,no.3,416(2016)].

\bibitem{Laughlin:1988fs}
R.~B. Laughlin, ``{The Relationship between high temperature superconductivity
  and the fractional quantum Hall effect},'' {\em Science} {\bf 242} (1988)
525--533.

\bibitem{Chen:1989xs}
Y.-H. Chen, F.~Wilczek, E.~Witten, and B.~I. Halperin, ``{On Anyon
  Superconductivity},'' {\em Int. J. Mod. Phys.} {\bf B3} (1989)
1001.

\bibitem{Fetter:1991ju}
A.~L. Fetter, R.~B. Laughlin, and C.~B. Hanna, ``{Anyons and superconductivity:
  Random phase approximation},'' {\em Int. J. Mod. Phys.} {\bf B5} (1991)
2751--2790.


\end{thebibliography}
\end{document}